\documentclass[aps,notitlepage,preprintnumbers,onecolumn,amsmath,amssymb,floatfix,superscriptaddress,pra]{revtex4-1}

\pdfoutput=1 
\usepackage[utf8]{inputenc}

\usepackage{amsmath}
\usepackage{amsfonts}
\usepackage{braket}
\usepackage{graphicx}
\usepackage{tensor}

\usepackage[usenames,dvipsnames,svgnames,table]{xcolor}
\usepackage[colorlinks=true,linkcolor=blue, citecolor=blue, bookmarks]{hyperref}

\newcommand{\be}{\begin{equation}}
\newcommand{\ee}{\end{equation}}
\newcommand{\bea}{\begin{eqnarray}}
\newcommand{\eea}{\end{eqnarray}}
\newcommand{\eps}{\varepsilon}

\begin{document}
\title{Entanglement dynamics of thermofield double states in integrable models}
\author{Gianluca Lagnese} 
\affiliation{SISSA and INFN, via Bonomea 265, 34136 Trieste, Italy}
\author{Pasquale Calabrese}
\affiliation{SISSA and INFN, via Bonomea 265, 34136 Trieste, Italy}
\affiliation{International Centre for Theoretical Physics (ICTP), Strada Costiera 11, 34151 Trieste, Italy}
\author{Lorenzo Piroli}
\affiliation{Philippe Meyer Institute, Physics Department, \'Ecole Normale Supérieure (ENS), Université PSL, 24 rue Lhomond, F-75231 Paris, France}

\begin{abstract}
We study the entanglement dynamics of thermofield double (TFD) states in integrable spin chains and quantum field theories. We show that, for a natural choice of the Hamiltonian eigenbasis, the TFD evolution may be interpreted as a quantum quench from an initial state which is low-entangled in the real-space representation and displays a simple quasiparticle structure. Based on a semiclassical picture analogous to the one developed for standard quantum quenches, we conjecture a formula for the entanglement dynamics, which is valid for both discrete and continuous integrable field theories, and expected to be exact in the scaling limit of large space and time scales. We test our conjecture in two prototypical examples of integrable spin chains, where numerical tests are possible. First, in the XY-model, we compare our predictions with exact results obtained by mapping the system to free fermions, finding excellent agreement. Second, we test our conjecture in the interacting XXZ Heisenberg model, against numerical iTEBD calculations. For the latter, we generally find good agreement, although, for some range of the system parameters and within the accessible simulation times, some small discrepancies are visible, which we attribute to finite-time effects.
\end{abstract}

\maketitle


\section{Introduction}
\label{sec:intro}

The \emph{thermofield double} (TFD) state is defined on two identical copies (or replicas) of a quantum many-body system or field theory, usually denoted by ``left'' and ``right'', and takes the form~\cite{israel1976thermo,maldacena2003eternal}
\be\label{eq:TFD_tzero}
\left|\operatorname{TFD}\right\rangle=\frac{1}{\sqrt{Z_{\beta}}} \sum_{n} e^{-\beta E_{n} / 2}\left|E_{n}\right\rangle_{L} \left|E_{n} \right\rangle_{R}\,,
\ee
where $E_{n}$ and $\ket{E_n}$ are the eigenvalues and eigenstates of the Hamiltonian $H$, $\beta$ is the inverse temperature, while $Z_\beta$ is the canonical partition function. The dynamics is given by evolving the left and right copies up to time $t_L$ and $t_R$, respectively, i.e.
\be\label{eq:TFD_time}
\left|\operatorname{TFD}\left(t_{L}, t_{R}\right)\right\rangle=\frac{1}{\sqrt{Z_{\beta}}} \sum_{n} e^{-\beta E_{n} / 2} e^{-i E_{n}\left(t_{L}+t_{R}\right)}\left|E_{n}\right\rangle_{L} \left|E_{n} \right\rangle_{R}\,.
\ee
The TFD state plays a very important role in the context of the AdS/CFT correspondence~\cite{maldacena1999large}, as it was proposed to be dual to an eternal black hole~\cite{israel1976thermo,maldacena2003eternal}. It provides a controlled setup to investigate various aspects of black-hole physics, and to explore new ideas inspired by quantum information theory, including questions related to quantum scrambling and chaos~\cite{shenker2014black,shenker2014multiple,roberts2015localized} and quantum complexity~\cite{stanford2014complexity,susskind2016entanglement,susskind2016addendum,brown2016complexity,brown2016holographic,yang2018complexity,yang2018comparison,chapman2019complexity,jiang2019circuit,doroudiani2020complexity,chapman2021charged}. 

The TFD state is also interesting and appears naturally in the study of non-relativistic quantum many-body systems, since it provides a \emph{purification} of the thermal Gibbs state. For instance, it is at the basis of several efficient tensor-network algorithms to compute thermal expectation values~\cite{schollwock2011density}.

When $t_L=t_R$, the TFD dynamics can be thought of as a quantum quench~\cite{calabrese2006time,calabrese2007quantum}. This is a protocol in which some well-defined initial state (for instance, the ground-state of some local Hamiltonian, or a simple low-entangled state) is left to evolve unitarily according to a local Hamiltonian, which should be sufficiently general with respect to the initial state (in particular, the latter should not be one of its eigenstates). In our case, the initial state is given by~\eqref{eq:TFD_tzero} and the post-quench Hamiltonian is
\be\label{eq:h_left_right}
H=H_L+H_R=H\otimes \openone+\openone\otimes H\,.
\ee
Given a subsystem made of two identical regions in the left and right spaces, a natural question pertains to the growth of the corresponding von Neumann entanglement entropy. On the one hand, this quantity is of interest in the context of holography~\cite{hartman2013time} and has already been investigated in many works, see for instance~\cite{hartman2013time, asplund2015entanglement, gu2017spread,mezei2017entanglement,mezei2018membrane,chapman2019complexity,jiang2019circuit,ghasemi2021odd,langlett2021rainbow,goto2021information}. On the other hand, when $\beta=0$, it coincides with the entanglement entropy of the evolution operator~\cite{zanardi2001entanglement}, which features in several recently-proposed measures of quantum scrambling and chaos for non-relativistic many-body systems, see e.g.~\cite{hosur2016chaos,zhou2017operator,sunderhauf2019quantum,schnaack2019tripartite,bertini2020scrambling,gong2021topological,styliaris2021information}. 

In general, it is very challenging to compute the growth of the TFD entanglement entropy after a quench: for short-range models, an exact calculation has been carried out only for non-interacting bosonic quantum field theories~\cite{chapman2019complexity}.  

For standard quenches, a powerful tool to capture the entanglement evolution beyond the non-interacting case is the \emph{quasiparticle picture}, originally proposed in $(1+1)$D conformal field theory (CFT)~\cite{calabrese2005evolution,calabrese2009entanglement,calabrese2016quantum}. Within this picture, one postulates that a homogeneous quench produces an extensive number of \emph{pairs} of quasiparticle excitations, which are responsible for propagating the entanglement throughout the system. 
It was later realized that the quasiparticle picture holds beyond quenches in CFT, and correctly describes the entanglement dynamics in one-dimensional ($1$D) systems with stable quasiparticle 
excitations \cite{c-20}. 
This was first shown for non-interacting spin-chains through a rigorous microscopic derivation~\cite{fagotti2008evolution}. 
More recently, it has been extended to \emph{interacting} integrable models~\cite{alba2017entanglement,alba2018entanglement} (see also~\cite{alba2019quantum,alba2019scrambling,modak2020entanglement,murciano2021quench,pbc-20,pbc-21, bc-20} for generalizations), inhomogeneous settings~\cite{bertini2018entanglement_evolution,alba2019evolution,alba2021generalized,mestyan-2020} and open quantum systems \cite{a-21,alba-d,carollo-d,alba-da}. 
These results are particularly interesting given the difficulty to compute the real-time dynamics in interacting many-body systems.
However, we should mention that the quasiparticle picture faces serious problems when applied to the R\'enyi entropies \cite{renyi,renyi-1,mestyan-2018,klobas2021entanglement}.
This is important because R\'enyi entropies are the entanglement proxies which have been measured in quench experiments with 
cold atoms and ion traps \cite{kaufman-2016,Elben2018,vek-21,brydges-2019}.

Contrary to the non-interacting case, the general validity of the quasiparticle picture has not yet been rigorously proven. Still, a significant amount of numerical evidence corroborating it has been collected in prototypical examples, including XXZ Heisenberg models~\cite{alba2017entanglement,alba2018entanglement} and $SU(3)$-invariant spin chains displaying multiple quasiparticle species~\cite{modak2019correlation}. In addition, it has been established analytically in certain quantum cellular automata mimicking interacting quasiparticle dynamics~\cite{klobas2021exact,klobas2021exact_II,klobas2021entanglement}. An important assumption of Refs.~\cite{alba2017entanglement,alba2018entanglement} is that the quench only produces pairs of quasiparticle excitations, which is known to be true for a class of low-entangled initial states~\cite{piroli2017what,pozsgay2019integrable}. In more general cases, contributions beyond quasiparticle pairs might become important and the formulas of~\cite{alba2017entanglement,alba2018entanglement} should be modified, as it was done explicitly in~\cite{bertini2018entanglement,bastianello2018spreading,bc-20b} for the case of non-interacting models.

Interestingly, the quasiparticle picture was recently tested to high precision against exact calculations for the TFD entanglement dynamics in non-interacting bosonic theories~\cite{chapman2019complexity}. This result is nontrivial: if we interpret the TFD state as the initial state for a quench problem in a two-replica space, it is not obvious, \emph{a priori}, that the picture developed for standard quenches should apply, as the setting is very different. Motivated by the results of~\cite{chapman2019complexity}, here we address a very simple question: \emph{can we extend the quasiparticle picture for the TFD entanglement dynamics to more general integrable models}? We stress that this question is interesting per se and goes beyond its possible connections to holography, as it also applies, for instance, to non-conformal theories and lattice models. 

In order to tackle this problem, it is first important to realize that the definition of the TFD state is not unique whenever the Hamiltonian displays spectral degeneracy: different choices for the energy eigenbasis lead to states with distinct physical properties. In this work, we will focus on one natural basis for which the TFD state is initially low-entangled in the real space representation. This is a natural choice to make contact with the standard theory of quantum quenches, where initial states are typically assumed to be low-entangled. 

With this definition, we show that the TFD dynamics can be interpreted as a quench where only pairs of quasiparticles are created, and conjecture a generalization of the standard picture for the dynamics of bipartite entanglement entropy. We test our conjecture in two prototypical examples of integrable spin chains. First, in the XY-model, we compare our predictions with exact results obtained by mapping the system to free fermions. Second, in the interacting XXZ Heisenberg model, we test our conjecture against numerical iTEBD calculations. 

The rest of this work is organized as follows. In Sec.~\ref{sec:quasiparticle_picture} we introduce the TFD state and discuss the choice of the Hamiltonian eigenbasis of interest in this work. We then review the quasiparticle picture for standard quenches, and present its generalization to the case of the TFD dynamics. In Sec.~\ref{sec:non-interacting_spin_chain} we study the case of non-interacting spin chains focusing on the XY model. By mapping the system to free fermions, we obtain an exact result for the TFD entanglement evolution, which allows us to perform a strict test of the quasiparticle picture. Next, Sec.~\ref{sec:XXZ} is devoted to the analysis of the interacting XXZ Heisenberg chain, where we compare our analytic predictions against numerical iTEBD calculations. Finally, we report our conclusions in Sec.~\ref{sec:conclusions}.

\section{The TFD state and the quasiparticle picture}
\label{sec:quasiparticle_picture}

\subsection{The TFD state and the Hamiltonian eigenbasis}

Let us begin by defining more precisely the quench protocol. For concreteness, we focus on a $1$D lattice model, with Hamiltonian $H$ and associated Hilbert space
\be
\mathcal{H}=h_1\otimes \ldots \otimes h_N\,,
\ee
where $h_j\simeq \mathbb{C}^{d}$, while $N$ is the system size. Following our discussion in Sec.~\ref{sec:intro}, we take two copies of the system, left ($L$) and right ($R$), so that the total Hilbert space is $\mathcal{K}=\mathcal{H}^L\otimes \mathcal{H}^R$. In the following, we will also denote by $h_j^{L/R}$ the left and right local Hilbert spaces. We study the quench protocol in which the system is initialized in the TFD state~\eqref{eq:TFD_tzero} and left to evolve according to the Hamiltonian~\eqref{eq:h_left_right}, so that the state at time $t/2$ after the quench is
\be\label{eq:TFD_quench}
\left|\operatorname{TFD}\left(t \right)\right\rangle=\frac{1}{\sqrt{Z_{\beta}}} \sum_{n} e^{-\beta E_{n} / 2} e^{-i E_{n}t}\left|E_{n}\right\rangle_{L} \left|E_{n} \right\rangle_{R}\,.
\ee

In the case the model displays spectral degeneracy, the state~\eqref{eq:TFD_quench} strongly depends on the choice of the Hamiltonian eigenbasis, which we thus need to specify in order to uniquely define the quench protocol. This choice is guided by the notion of locality that we set in the two-replica space: Interpreting the doubled Hilbert space as a lattice with local site $k_j=h^{L}_j\otimes h^{R}_{j}$, we require that the TFD state~\eqref{eq:TFD_tzero} is a low entangled state in this lattice. This choice is very natural, and allows us to make direct contact with the theory of quantum quenches, where initial states are typically assumed to be low entangled.

Let us show that there always exists a Hamiltonian eigenbasis for which the TFD state at $\beta=0$ is a \emph{product state} in the doubled lattice $\mathcal{K}$. Denoting by $\{\ket{\alpha}_j\}_{\alpha=1}^{d}$ a basis for $h_j$, we introduce the following maximally entangled state between $\mathcal{H}^{L}$ and $\mathcal{H}^{R}$
\be
\ket{\mathcal{I}}=\frac{1}{\sqrt{d^N}}\sum_{\{\alpha_n\}}\ket{\alpha_1, \ldots, \alpha_N}\otimes \ket{\alpha_1, \ldots, \alpha_N}\in \mathcal{K}\,,
\ee
where the sum is over all sequences $\{\alpha_1,\ldots, \alpha_N\}$ with $\alpha_k=1,\ldots d$. Importantly, $\ket{\mathcal{I}}$ can be written as a product state in $\mathcal{K}$, namely
\be\label{eq:product_infinite}
\ket{\mathcal{I}}=\bigotimes_{k=1}^{N}\left(\frac{1}{\sqrt{d}}\sum_{\alpha=1}^{d}\ket{\alpha}_{L,k}\otimes \ket{\alpha}_{R,k}\right)\,.
\ee
For the models considered in this work, it is easy to check that one can choose a local basis $\{\ket{\alpha}\}_{\alpha=1}^d$ such that
\be\label{eq:ham_condition}
\overline{\braket{\alpha_1, \ldots, \alpha_N|H|\alpha_1, \ldots, \alpha_N}}=\braket{\alpha_1, \ldots, \alpha_N|H|\alpha_1, \ldots, \alpha_N}\,,
\ee
where $\overline{(\cdot)}$ denotes complex conjugation (which is a basis-dependent operation). When this condition holds, the Hamiltonian then clearly displays time-reversal symmetry, cf. also our discussion in Sec.~\ref{sec:TFD_conjecture}. It follows that there exists a Hamiltonian eigenbasis $\{\ket{E_n}\}_n$ which satisfies
\be\label{eq:TFD_eigenbasis}
\overline{\braket{\alpha_1, \ldots, \alpha_N|E_n}}=\braket{\alpha_1, \ldots, \alpha_N|E_n}\,.
\ee
Therefore, denoting by $\mathcal{N}$ the matrix which implements the change of basis from $\{\ket{\{\alpha_n\}}\}$ to $\{\ket{E_n}\}$, we obtain that $\bra{\{k\}}\mathcal{N}\ket{\{j\}}=\overline{\bra{\{k\}}\mathcal{N}\ket{\{j\}}}$, and so
\be\label{eq:orthogonality}
\mathcal{N}\mathcal{N}^{T}=\mathcal{N}\mathcal{N}^{\dagger}=\openone\,.
\ee
Finally, using~\eqref{eq:orthogonality}, it is a straightforward to show that
\be\label{eq:real_space_representation_beta0}
\left|\operatorname{TFD}(0)\right\rangle=\frac{1}{d^{N/2}} \sum_{n} \left|E_{n}\right\rangle_{L} \left|E_{n} \right\rangle_{R}=\ket{\mathcal{I}}\,.
\ee
In the case where~\eqref{eq:ham_condition} does not hold, one may define an alternative TFD state by complex-conjugating the vectors in the second replica space. However, the integrable models considered in this work all display time-reversal symmetry, so we will not consider this case in the following. 

Eq.~\eqref{eq:real_space_representation_beta0} gives us the real-space representation of the infinite-temperature TFD state considered in this work. We note that there is still an ambiguity in its definition, since it depends on the choice of the basis of the local Hilbert space. However, different choices are now related by local unitary transformations, which do not modify the bipartite entanglement. The real-space representation of the finite-time and finite-temperature TFD state is thus simply
\be\label{eq:real_space_representation}
\left|\operatorname{TFD}(t)\right\rangle=\exp\left[-i\frac{t}{2} \left(H_L\otimes \openone + \openone \otimes H_R \right)\right]\exp\left[-\frac{\beta}{4}\left(H_L\otimes \openone + \openone \otimes H_R \right)\right]\ket{\mathcal{I}}\,.
\ee

We consider now a bipartition of the two-replica space into the region $A_{\ell}$ and its complement $A^c_{\ell}$, where $A_{\ell}$ contains the first $\ell$ sites of both the left and right lattices. The associated Hilbert space is
\be
\mathcal{K}_{A}=(h^L_1\otimes h^R_1)\otimes  (h^L_2\otimes h^R_2)\otimes\ldots\otimes (h^L_\ell\otimes h^R_\ell)\,.
\ee
We will be interested in the dynamics of the entanglement between $A_\ell$ and $A^c_\ell$, which can be quantified by the von Neumann entanglement entropy
\be\label{eq:vn_entanglement}
S_{A_\ell}(t)=- {\rm tr}\rho_{A_\ell}(t)\ln\rho_{A_\ell}(t)\,,
\ee
where $\rho_{A_\ell}(t)$ is the density matrix at time $t$ reduced to the subsystem $A_\ell$, i.e.
\be
\rho_{A_\ell}(t)= {\rm tr}_{\!A^c_\ell} \ket{\operatorname{TFD}\left(t \right)}\!\bra{\operatorname{TFD}\left(t \right)}\,.
\ee
Note that, because of~\eqref{eq:real_space_representation}, $S_{A_\ell}(0)=0$ for $\beta=0$. 

We will formulate a conjecture for the von Neumann entanglement entropy~\eqref{eq:vn_entanglement} in the scaling limit $\ell, t\to\infty$, where the ratio $\ell/t$ is kept constant. Our conjecture is based on the quasiparticle picture for standard quenches, which we briefly review in the next subsection.
 
\subsection{The quasiparticle picture}

As mentioned, the quasiparticle picture was originally introduced in the context of homogeneous quenches in $(1+1)$D CFT~\cite{calabrese2005evolution}. In this setting, the system is initialized in a pure state $\ket{\Psi}$, for instance the ground-state of a local Hamiltonian $H_0$, and left to evolve under a post-quench local Hamiltonian $H$ (which is assumed to be translation-invariant). The quasiparticle picture provides a very simple interpretation (and quantitative description) of the post-quench entanglement dynamics, as we briefly explain now (see the reviews~\cite{calabrese2009entanglement,calabrese2016quantum} for a comprehensive treatment). 

As a starting point, one postulates that the homogeneous quench produces everywhere an extensive number of uncorrelated pairs of entangled quasiparticles with opposite momenta. After the quench, the quasiparticles spread through the system, and cause distant regions to be entangled. At a given time $t$, the entanglement between two regions, $A$ and $B$, is proportional to the number of pairs with one quasiparticle in $A$ and the other in $B$. In CFT, the quasiparticles all propagate with the same velocity $v$, leading to the very simple equation
\begin{equation}
\label{eq:cft}
S_{\ell} (t)=4v t s\Theta(\ell-2v t)+\ell s\Theta(2v t-\ell)\,,
\end{equation}
where $\Theta$ is the Heaviside theta function, we used the short-hand notation $S_{\ell}=S_{A_\ell}$ and $\ell$ is the subsystem size. Here $s$ is the entanglement-entropy density carried by each pair of quasiparticles. Eq.~\eqref{eq:cft} has been derived assuming that the infinite-system size limit $N\to\infty$ is taken first.

Recently, this picture has been generalized to the case of interacting integrable systems~\cite{alba2017entanglement, alba2018entanglement}. As a defining feature, integrable models are characterized by the fact that the entire spectrum of their Hamiltonian can be described in terms of \emph{stable} quasiparticles. In analogy with the non-interacting case, they can be parametrized by quasimomenta, or \emph{rapidities} $\{\lambda_j\}_j$, and display a non-trivial dispersion relation. The latter has to be taken into account within the quasiparticle picture, implying that distinct quasiparticles propagate with different velocities. This leads to a simple modification of Eq.~\eqref{eq:cft}, where one sums over the contributions coming from quasiparticles with different rapidities $\lambda$, namely~\cite{alba2017entanglement}
\be\label{eq:quasiparticle}
S_{\ell}(t) = \sum_n\left[2 t \int_{2|v_n| t<\ell} \mathrm{~d} \lambda|v_n(\lambda)| s_n(\lambda)+\ell \int_{2|v_n| t>\ell} \mathrm{~d} \lambda s_n(\lambda)\right]\,.
\ee
Here, $v(\lambda)$ and $s(\lambda)$ denote the velocity and density of entanglement entropy of the quasiparticles with rapidity $\lambda$, respectively. We also introduced an additional index $n$, which distinguishes between possible different types of quasiparticles or bound states thereof.

It was argued in Ref.~\cite{alba2017entanglement} that $v_n(\lambda)$ and $s_n(\lambda)$ can be related to the thermodynamic properties of the stationary state emerging at large times after the quench, as described by the Generalized Gibbs Ensemble (GGE)~\cite{vidmar2016generalized,essler2016quench}. In particular, $v_n(\lambda)$ was identified with the velocity of the elementary excitations, and $s_n(\lambda)$ with the corresponding thermodynamic, or Yang-Yang, entropy~\cite{takahashi2005thermodynamics}. As a consequence, $v_n(\lambda)$ and $s_n(\lambda)$ can be computed exactly (at least for simple initial states), and \eqref{eq:quasiparticle} gives us a fully quantitative prediction with no free parameter to be fixed~\cite{alba2017entanglement}.

An important assumption underlying Eq.~\eqref{eq:quasiparticle} is that the quench only produces pairs of quasiparticles. At the microscopic level, this can be justified for a class of low-entangled \emph{integrable} initial states~\cite{piroli2017what,pozsgay2019integrable} (see also~\cite{delfino2014quantum,delfino2017theory} for related discussions). However, in more general cases it is possible that the quench also produces higher $n$-tuples of correlated quasiparticles. An explicit example in a non-interacting model is reported in Ref.~\cite{bertini2018entanglement}, where the authors found fine-tuned families of initial states for which the quench does not produce pairs but only higher $n$-tuples of quasiparticles. In this case, the entanglement dynamics is still captured by a semiclassical picture, but Eq.~\eqref{eq:quasiparticle} has to be modified~\cite{bertini2018entanglement}.

\subsection{The TFD quasiparticle content and entanglement dynamics: a conjecture}
\label{sec:TFD_conjecture}

From the above discussions, it is not obvious that the standard quasiparticle picture should apply to the TFD dynamics: while the spectrum of the two-replica Hamiltonian~\eqref{eq:h_left_right} is still characterized in terms of stable quasiparticles, the TFD state is a very complicated superposition of eigenstates, and one could wonder whether higher $n$-tuples could contribute. Here we show that, for Bethe Ansatz solvable models, the TFD state~\eqref{eq:real_space_representation} admits a representation from which the assumption of pairs of quasiparticles appears to be fully justified.

In the following discussion, we will consider an integrable spin chain (or field theory) with periodic boundary conditions, whose eigenstates are parametrized by the sets of quasiparticle rapidities $\{\lambda_j\}$. In addition, we will assume that the real-space representation of the eigenstates $|E\left(\{\lambda_j\} \right)\rangle$ is such that 
\be
\overline{ \braket{\{j\}| E\left(\{\lambda_j\} \right)} }= \braket{\{j\}| E\left(\{-\lambda_j\} \right)}\,.
\label{eq:conjugate_property}
\ee
This property can be verified at the level of the eigenfunctions in all integrable models which will be considered in this work~\cite{korepin1997quantum}. In fact, this relation has a physical meaning: provided that the Hamiltonian is real in the computational basis $\ket{\{\alpha_k\}}$, complex conjugation corresponds to time inversion, which has the effect of flipping the sign of the quasiparticle momenta. 

Since eigenstates with opposite sets of rapidities have the same energy, we can define the new Hamiltonian eigenstates
\bea\label{eq:e_plusminus}
\ket{\mathcal{E}^+(\{\lambda_j\})}&=&\frac{1}{\sqrt{2}}\ket{E\left(\{\lambda_j\} \right)}+\ket{E\left(\{-\lambda_j\} \right)}\,,\\
\ket{\mathcal{E}^-(\{\lambda_j\})}&=&\frac{i}{\sqrt{2}}\left(\ket{E\left(\{\lambda_j\} \right)}-\ket{E\left(\{-\lambda_j\} \right)}\right)\,.
\eea
Now, the eigenbasis $\{\ket{\mathcal{E}^{\pm }(\{\lambda_j\})}\}$ satisfies~\eqref{eq:TFD_eigenbasis}, and so
\be
\ket{\mathcal{I}}\propto \sum_{\alpha=\pm } \sum_{\{\lambda_j\}} \ket{\mathcal{E}^{\alpha}(\{\lambda_j\})}\otimes \ket{\mathcal{E}^{\alpha}(\{\lambda_j\})}\,.
\ee
Using~\eqref{eq:e_plusminus}, we finally obtain
\be\label{eq:pair_decomposition}
\ket{\mathcal{I}}=\frac{1}{d^{N/2}}\sum_{\{\lambda_j\}} \ket{E(\{\lambda_j\})}\otimes \ket{E(\{-\lambda_j\})}\,.
\ee
The interpretation of $\ket{\mathcal{I}}$ in terms of quasiparticles is now clear: it is a superposition of eigenstates, in which any quasiparticle moving in the first replica space with rapidity $\lambda_j$ is paired to one moving in the second replica space with opposite rapidity $-\lambda_j$. 

Eq.~\eqref{eq:pair_decomposition} provides a basis for the application of the standard quasiparticle picture to the TFD dynamics. Based on the latter, we conjecture that Eq.~\eqref{eq:quasiparticle} describes \emph{exactly} the TFD entanglement dynamics in the scaling limit of large $t$ and $\ell$, with the ratio $t/\ell$ kept constant.

In order to give Eq.~\eqref{eq:quasiparticle} predictive power, we need to specify the functions $s_n(\lambda)=s_{(n)}^{\rm TFD}(\lambda)$ and $v_n(\lambda)=v_{(n)}^{\rm TFD}(\lambda)$. Following the logic of Refs.~\cite{alba2017entanglement,alba2018entanglement}, they are determined by the stationary state (GGE) emerging at large times after a quench from the TFD state. Since the latter is a purification of the Gibbs state, the expectation value of all the local conserved quantities coincide with the thermal one, so that the GGE is simply the Gibbs ensemble with inverse temperature $\beta$. Thus, we arrive at the identification (dropping the index $n$)
\begin{subequations}\label{eq:conjecture}
	\begin{align}
		v^{\rm TFD}(\lambda)&=v^{ \beta}(\lambda)\,,\label{eq:v_thermal}\\
		s^{\rm TFD}(\lambda)&=2s^{ \beta}(\lambda)\,,\label{eq:s_thermal}
	\end{align}
\end{subequations}
where $v^{ \beta}(\lambda)$ and $s^{ \beta}(\lambda)$ are the thermal velocity of excitations and thermodynamic entropy at inverse temperature $\beta$, respectively. The factor of $2$ in Eq.~\eqref{eq:s_thermal} follows from the fact that the local dimension in the doubled space is the square of the original one. The quasiparticle prediction consisting of Eqs.~\eqref{eq:quasiparticle} and \eqref{eq:conjecture} first appeared in Ref.~\cite{chapman2019complexity}, where it has been tested against analytic calculations in free scalar quantum field theories. 

We stress that the validity of the quasiparticle picture formulated above is a conjecture: as in the case of standard quenches~\cite{alba2017entanglement,alba2018entanglement}, it is highly non-trivial to derive rigorously predictions in the scaling limit starting from the microscopic theory, even if the explicit spectral decomposition of the initial state is known, cf. Eq.~\eqref{eq:pair_decomposition}. Although we expect that in the non-interacting case a rigorous derivation could be carried out by generalizing Ref.~\cite{fagotti2008evolution}, at the moment a proof in the presence of interactions appears to be out of reach. In the next sections, we will provide strong evidence of its validity by comparison against analytic and numerical calculations in concrete integrable spin chains.

\section{The XY model}
\label{sec:non-interacting_spin_chain}

We begin our analysis of the TFD entanglement dynamics by focusing on a prototypical example of a non-interacting spin chain, the so-called XY model, whose Hamiltonian reads
\begin{equation}\label{eq:xxham}
	H_{XY}(\gamma,h)=-\frac{1}{2} \sum_{j=1}^{N}\left[\left(\frac{1+\gamma}{2}\right) \sigma_{j}^{x} \sigma_{j+1}^{x}+\left(\frac{1-\gamma}{2}\right) \sigma_{j}^{y} \sigma_{j+1}^{y}+h \sigma_{j}^{z}\right]\,,
\end{equation}
where $\sigma^\alpha_j$ are the Pauli matrices, and periodic boundary conditions are assumed. Here $\gamma$ and $h$ are the anisotropy parameter and magnetic field, respectively. The spectrum of this model can be obtained exactly by means a Jordan-Wigner (JW) transformation. Introducing the fermionic modes
\begin{align}
c_j&=\left(\bigotimes_{k=1}^{j-1}\sigma_k^{z}\right)\sigma_j^{-}\,, \qquad  c^\dagger_j=\left(\bigotimes_{k=1}^{j-1}\sigma_k^{z}\right)\sigma_j^{+}\,,
\end{align}
where $\sigma^\pm=(\sigma^x\pm i\sigma^y)/2$, the Hamiltonian~\eqref{eq:xxham} is mapped onto
\begin{equation}
\begin{aligned}
H=-& \frac{1}{2} \sum_{j=1}^{N-1}\left(c_{j}^{\dagger} c_{j+1}+c_{j+1}^{\dagger} c_{j}+\gamma c_{j}^{\dagger} c_{j+1}^{\dagger}+\gamma c_{j+1} c_{j}\right)+h \sum_{j=1}^{N} c_{j}^{\dagger} c_{j}-\frac{h N}{2} \\
&+\frac{P}{2}\left(c_{N}^{\dagger} c_{1}+c_{1}^{\dagger} c_{N}+\gamma c_{N}^{\dagger} c_{1}^{\dagger}+\gamma c_{1} c_{N}\right)\,.
\end{aligned}
\end{equation}
Here $P= \prod_{j=1}^{N}\left(1-2 c_{j}^{\dagger} c_{j}\right)$ is the parity operator, which determines periodic (antiperiodic) boundary conditions in the sector of odd (even) fermionic numbers. Since we will be interested in the scaling limit of the entanglement of large subsystem sizes, we can neglect boundary effects, and focus on the fermionic Hamiltonian with periodic boundary conditions~\cite{Note1}, i.e.
\begin{equation}\label{eq:quadratic_fermion}
\begin{aligned}
	H=-& \frac{1}{2} \sum_{j=1}^{N}\left(c_{j}^{\dagger} c_{j+1}+c_{j+1}^{\dagger} c_{j}+\gamma c_{j}^{\dagger} c_{j+1}^{\dagger}+\gamma c_{j+1} c_{j}\right)+h \sum_{j=1}^{N} c_{j}^{\dagger} c_{j}-\frac{h N}{2} \,.
\end{aligned}
\end{equation}

In order to obtain a mapping to free fermions in the TFD setting, one needs to apply the JW transformation to both replicas independently. However, this procedure leads to a subtlety which needs to be taken into account. Indeed, since operators in the two spaces commute, applying the JW transformation to the two replicas independently yields mixed commutation relations: by construction, given two operators obtained by the JW transformation, they commute if they act on distinct replicas, and anti-commute otherwise. On the other hand, in order to map the spin system onto a truly fermionic one, all the transformed operators should anti-commute. Luckily, there is a simple way to get around this problem, which consists in a redefinition of the fermionic operators. Since this is a rather technical point, we discuss it in Appendix~\ref{sec:TFD_JW}. 

Crucially, the JW transformation maps the space of the first $\ell$ spins onto that of the first $\ell$ fermions, for all $\ell<N$. As a consequence, the bipartite entanglement entropy of the original spin chain can be obtained from the corresponding fermionic system~\cite{vidal2003entanglement}. Putting all together, we are left with the problem of computing the bipartite TFD entanglement dynamics for a quadratic fermionic Hamiltonian. 

This problem is now analogous to that treated in Ref.~\cite{chapman2019complexity}, where quadratic \emph{bosonic} field theories were considered. There, the TFD entanglement dynamics was computed exactly using that the TFD state is Gaussian, i.e. it satisfies Wick's theorem. This allows one to express its entanglement entropy in terms of the covariance matrix~\cite{vidal2003entanglement}, which, in turn, can be computed efficiently at any time. This logic can be followed without modifications also for fermionic degrees of freedom. In the following, we carry out this program explicitly for the model~\eqref{eq:quadratic_fermion}. 

\begin{figure}
	\centering
	\includegraphics[width=1
	\textwidth]{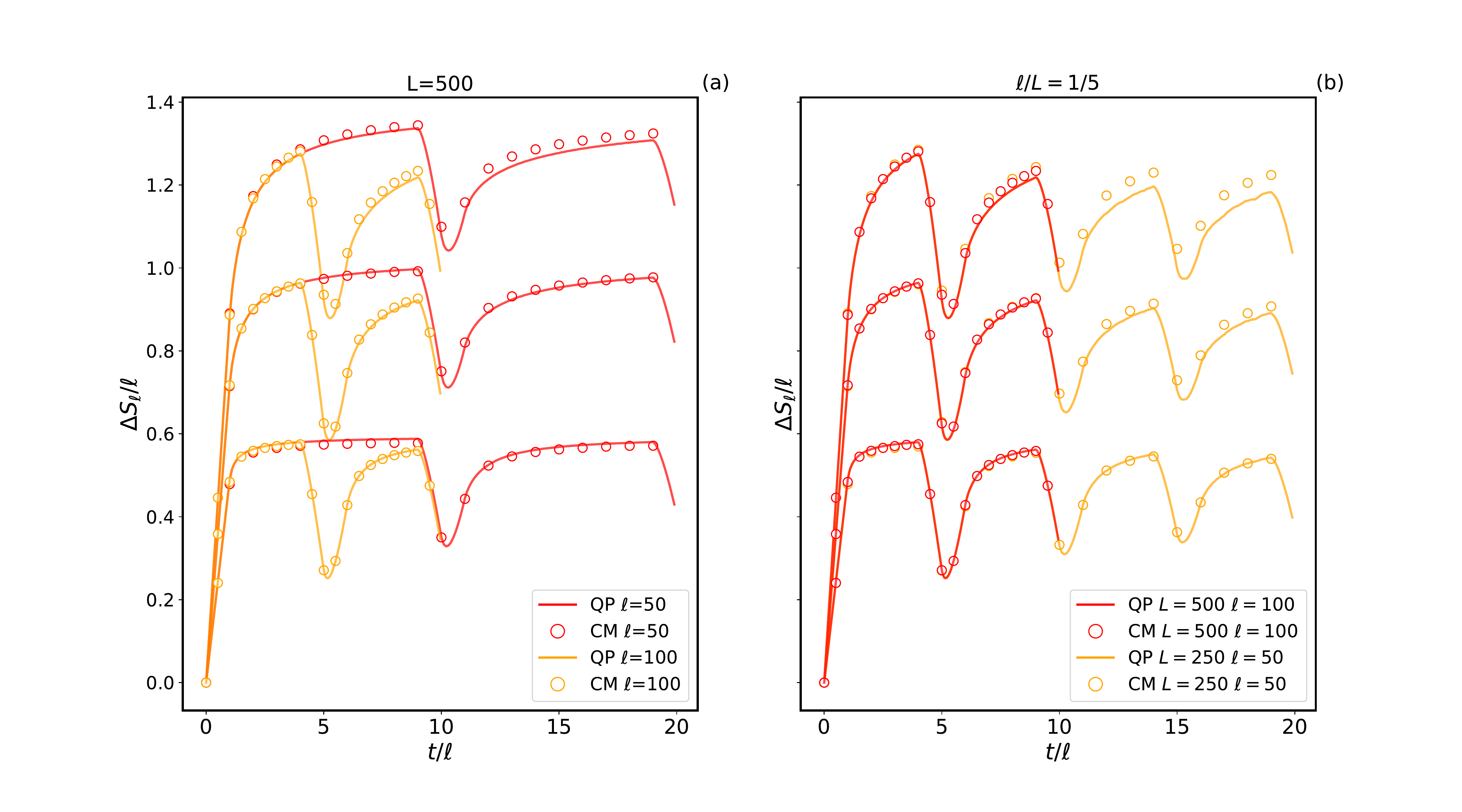}
	\caption{Comparison between the exact entanglement dynamics and quasiparticle predictions in the XY spin chain. The system parameters are $\gamma=h=0$, while in both plots the reverse temperature is $\beta=0,2,4$, increasing from the top to the bottom. In the left panel, the system size is fixed to $L=500$, and different subsystem sizes are considered. In the right panel, we fix the ratio $\ell/L$, and explore the dependence on the system size $L$. In both figures, we plot $\Delta S_\ell(t)/\ell=[S_\ell(t)-S_\ell(0)]/\ell$, where $S_\ell(0)$ is the bipartite entanglement entropy in the initial state.
	}\label{fig:XX_dynamics}
\end{figure}

As a first step, we recall that the Hamiltonian~\eqref{eq:quadratic_fermion} may be brought to a diagonal form by combining the Fourier transform with the Bogoliubov rotation~\cite{franchini2017introduction}, yielding
\be\label{eq:diagonal_free_fermion}
H= \sum_{q =0}^{N-1} \varepsilon\left(k_q\right)\left\{b_{q}^{\dagger} b_{q}-\frac{N}{2}\right\}\,,
\ee
where $k_q=\frac{2 \pi}{N} q$, and
\be\label{eq:energy_ff}
\varepsilon(\alpha) \equiv \sqrt{(h-\cos \alpha)^{2}+\gamma^{2} \sin ^{2} \alpha}\,.
\ee
Here $b_j=U_{j,k}c_k$, where $U_{j,k}$ is a suitably defined unitary matrix, whose explicit expression is given in Appendix~\ref{sec:details_TFD_free}. In this basis, the TFD~\eqref{eq:pair_decomposition}  reads
\be\label{eq:TFD_fermionic}
\ket{\rm TFD(t,\beta)}=\frac{1}{2^{N/2}}\sum_{\{q_n\}} \exp\left\{-it\sum_j[\varepsilon(k_{q_j})-\frac{\beta}{2}\varepsilon(k_{q_j})]\right\} (b^{L\dagger}_{q_1} b^{R\dagger}_{-q_1})\ldots (b^{L\dagger}_{q_n} b^{R\dagger}_{-q_n})\ket{0}\,,
\ee
where $\ket{0}$ is the vacuum associated with the modes $b_j$, i.e. $b_j\ket{0}=0$ for all $b_j$.  Note that $\ket{\rm TFD(0,0)}$ is a product state of Bell pairs, with the modes $b^L_{q_k}$, $b^R_{q_{-k}}$ being maximally entangled. 

The covariance matrix of~\eqref{eq:TFD_fermionic} is block-diagonal, and thus very easy to compute, as we detail in Appendix~\ref{sec:details_TFD_free} (similar calculations were performed in Ref.~\cite{jiang2019circuit}). Specifically, introducing the Majorana modes
\begin{align}\label{eq:majoranas_k}
	\psi^{L}_{2k} &= \frac{1}{\sqrt{2}} (b_k^{L\dag} + b^{L}_k)\,,\qquad 
	\psi^{L}_{2k+1} = \frac{i}{\sqrt{2}} (b_k^{L\dag} - b^{L}_k)\,, \\
	\psi^{R}_{2k} &= \frac{1}{\sqrt{2}} (b_{-k}^{R\dag} + b^{R}_{-k})\,, \qquad
	\psi^{R}_{2k+1} = \frac{i}{\sqrt{2}} (b_{-k}^{R\dag} - b^{R}_{-k}) \,,
\end{align}
we get the expectation value of $\psi^{\alpha}_{r_j} \psi^{\beta}_{s_j}$ in the state~\eqref{eq:TFD_fermionic}, which reads
\begin{equation}\label{eq:omega_k}
\begin{pmatrix}
		0 & \sin[2 \varphi(k_j)] \sin[\eps(k_j) t] & \cos[2 \varphi(k_j)] & \sin[2 \varphi(k_j)]  \cos[\eps(k_j) t] &  \\
		-\sin[2 \varphi(k_j)] \sin[\eps(k_j) t] & 0 & -\sin[2 \varphi(k_j)]  \cos[\eps(k_j) t] &  \cos[2 \varphi(k_j)] \\
		-\cos[2 \varphi(k_j)] & \sin[2 \varphi(k_j)]  \cos[\eps(k_j) t] & 0 &  -\sin[2 \varphi(k_j)] \sin[\eps(k_j) t] \\
		-\sin[2 \varphi(k_j)]  \cos[\eps(k) t] & -\cos[2 \varphi(k_j)] &  \sin[2 \varphi(k_j)] \sin[\eps(k_j) t] & 0 
	\end{pmatrix},
\end{equation}
with
\be
\varphi(k)=\arctan( e^{-\beta \eps (k)/2})\,,
\ee
where $\alpha,\beta=L,R$, $r_j,s_j=2j, 2j+1$ and the Majorana modes are ordered as $(\psi^{L}_{2j}, \psi^{R}_{2j}, \psi^{L}_{2j+1},\psi^{R}_{2j+1})$. Finally, setting 
\begin{align}\label{eq:initial_majorana}
	\chi^{\alpha}_{2k} &= \frac{1}{\sqrt{2}} (c_k^{\alpha\dag} + c^{\alpha}_k)\,,\qquad 
	\chi^{\alpha}_{2k+1} = \frac{i}{\sqrt{2}} (c_k^{\alpha\dag} - c^{\alpha}_k)\,,
\end{align}
and expressing the modes $\{c_j\}$ in terms of $\{b_j\}$, we obtain the covariance matrix
\be\label{eq:covariance_matrix}
\Gamma^{\alpha,\beta}_{j,k}(t,\beta)=\braket{{\rm TFD(t,\beta)}|\chi^{\alpha}_j \chi^{\beta}_k |{\rm TFD(t,\beta)}}\,.
\ee
Note that the definition of the left/right Majorana modes~\eqref{eq:majoranas_k} involve opposite quasi-momenta, which must be taken into account when expressing $\{c_j\}$ in terms of $\{b_j\}$ [cf. Appendix~\ref{sec:details_TFD_free} for details].

\begin{figure}
	\centering
	\includegraphics[width=1
	\textwidth]{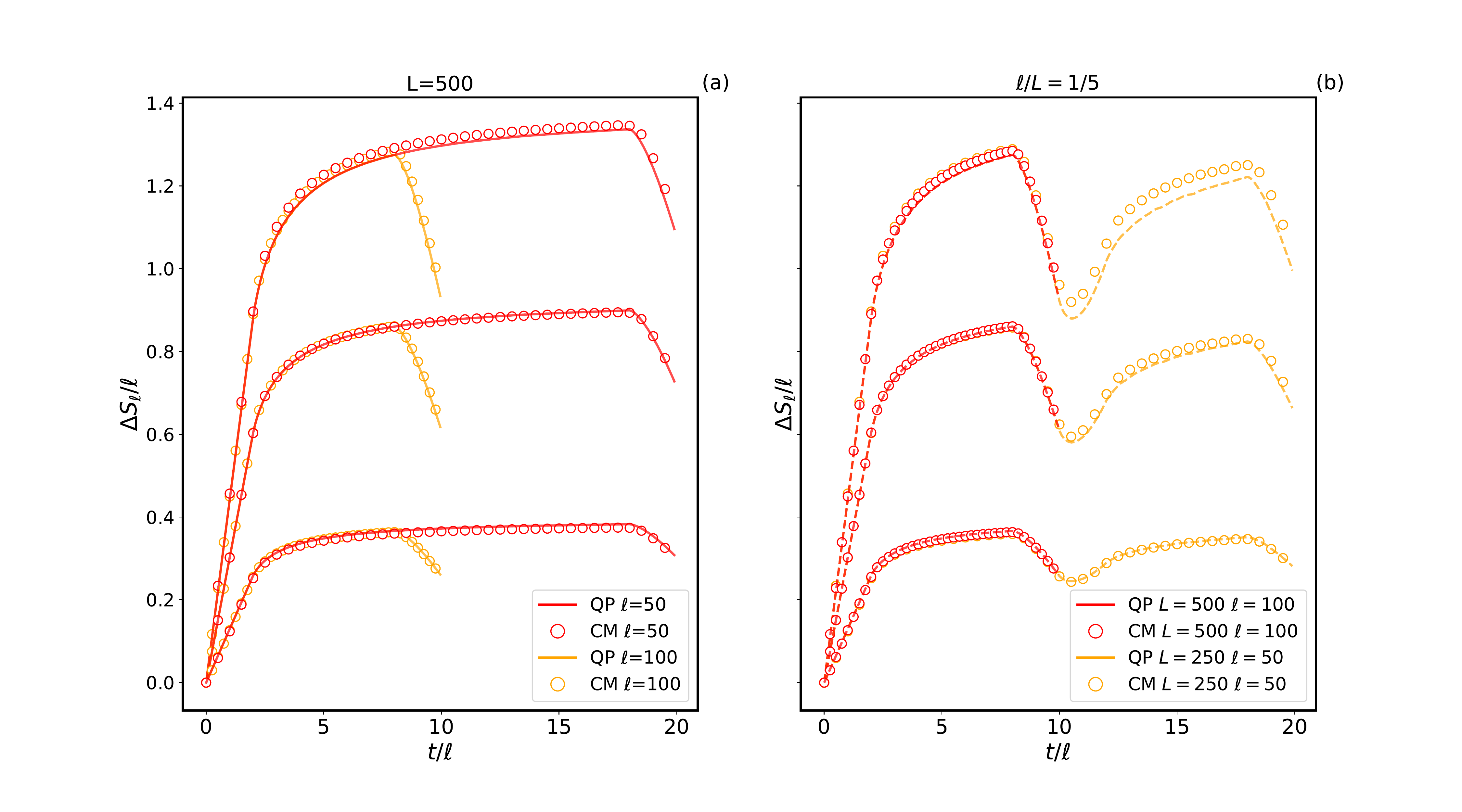}
	\caption{Same as Fig.~\ref{fig:XX_dynamics}. The system parameters are $\gamma=0.5$, $h=0$, while in both plots the reverse temperature is $\beta=0,2,4$, increasing from the top to the bottom}
	\label{fig:XY_dynamics}
\end{figure}

The covariance matrix~\eqref{eq:covariance_matrix} now gives us direct access to the bipartite von Neumann entanglement entropy $S_{\ell}(t)$~\cite{vidal2003entanglement}: denoting by $\Gamma_\ell(t)$ the covariance matrix restricted to the first $\ell$ sites, we have
\be\label{eq:quadratic_entanglement}
S_{\ell}(t)=-\operatorname{tr}\left[\frac{1+i \Gamma_{\ell}(t)}{2} \ln \frac{1+i \Gamma_{\ell}(t)}{2}\right]\,.
\ee 
This formula can be evaluated in a numerically exact and efficient way for large system sizes and times, and for arbitrary values of the Hamiltonian parameters~\eqref{eq:xxham}. It thus provides the possibility of a very stringent test for our analytic predictions.

Eq.~\eqref{eq:conjecture}  has been derived assuming that the limit $N\to\infty$ is taken first, i.e. for $\ell/N\to 0$. On the other hand, when evaluating~\eqref{eq:quadratic_entanglement}, finite-$\ell$ effects are visible even for relatively large sizes. Therefore, in order to test the quasiparticle picture and following Ref.~\cite{chapman2019complexity}, it is first convenient to adapt the formula~\eqref{eq:conjecture} to a ring geometry where the ratio $\ell/N$ is kept constant. This also allows us to explore the interesting revivals of the entanglement dynamics. By the usual argument, counting the number of pairs with one quasiparticle in $A_\ell$ and the other in $A_\ell^{c}$, and taking into account periodic boundary conditions, we arrive at the formula
\be\label{eq:quasiparticles_ring}
S_\ell(t)/\ell=  \int_{-\pi}^{\pi}{\rm d}k \left\{
\begin{array}{ll}
	(N/\ell) s^{\rm TFD}(k)  {\rm frac} \left(\frac{v(k) t}{N}\right) &\qquad \mbox{if }  N{\rm frac}(\frac{v t}{N})<\ell\,, \\
	s^{\rm TFD}(k)& \qquad \mbox{if }   \ell \leq N {\rm frac}(\frac{v(k) t}{N})<N-\ell\,, \\
	(N/\ell)  s^{\rm TFD}(k) \left[1-{\rm frac}\left(\frac{v(k) t}{N}\right) \right]& \qquad \mbox{if }  N-\ell <N {\rm frac}\left(\frac{v(k) t}{N}\right)\,,
\end{array}
\right.
\ee
where ${\rm frac}(\cdot)$ denotes the fractional part, e.g. ${\rm frac}(3/2)=1/2$. Since the system is non-interacting, the velocity and entropy contributions of the quasiparticles are not ``dressed'', and can be immediately read off from the diagonal form of the Hamiltonian~\eqref{eq:diagonal_free_fermion}, yielding
\begin{align}
v^{\rm TFD}(k)&=\frac{d\varepsilon(k)}{d k}\,,
\label{eq:thermalentropydensity}\\
s^{\rm TFD}(k)&=2\left[\frac{\beta  \varepsilon(k)}{1+e^{\beta  \varepsilon(k)}}+\log(1+e^{-\beta   \varepsilon(k)} )\right]\,,
\end{align}
where $\varepsilon(k)$ is given in~\eqref{eq:energy_ff}, while the extra factor accounts for the doubling of the local Hilbert space dimension. 

We have evaluated Eqs.~\eqref{eq:quadratic_entanglement} and ~\eqref{eq:quasiparticles_ring} for large systems sizes and different values of the system parameters, and systematically compared the results. Examples of our numerical data is given in  Figs.~\ref{fig:XX_dynamics} and~\ref{fig:XY_dynamics}. We find that, at short times and high temperatures, the agreement is extremely good even for relatively small system sizes. As the temperature is lowered and wider time intervals are considered, increasing system sizes are needed in order to observe the same accuracy in the quasiparticle predictions. We also see that the entanglement revivals are perfectly captured by the analytic formula~\eqref{eq:quasiparticles_ring}. In general, we have found excellent agreement between the exact entanglement dynamics and the quasiparticle prediction, giving us very strong evidence for the validity of the conjecture in the non-interacting case. 

\section{The XXZ Heisenberg spin chain}
\label{sec:XXZ}

We move on to test our conjecture in genuinely interacting integrable systems and consider the prototypical example of the XXZ Heisenberg model~\cite{korepin1997quantum}
\be\label{eq:xxz_hamiltonian}
H_{\mathrm{XXZ}}=\frac{1}{4}\sum_{j=1}^{N}\left[\sigma_{j}^{x} \sigma_{j+1}^{x}+\sigma_{j}^{y} \sigma_{j+1}^{y}+\Delta\left(\sigma_{j}^{z} \sigma_{j+1}^{z}-\frac{1}{4}\right)\right]\,,
\ee
where $\sigma^{\alpha}_j$ are Pauli matrices acting on the local space $h_j\simeq \mathbb{C}^{2}$, while $\Delta$ is the anisotropy parameter.

Although the model is integrable, interactions make it notoriously hard to analyze its out-of-equilibrium dynamics analytically~\cite{calabrese2016introduction}. For instance, it is not known how to compute the evolution of any non-trivial physical quantity following a quantum quench (with the only exception being the so-called Loschmidt echo~\cite{piroli2017from, piroli2018non}). Accordingly, in order to obtain an exact description of the TFD dynamics, we need to rely on numerical methods of general validity.

Since we are interested in the limit of large system sizes, we employ tensor-network (TN) methods~\cite{schollwock2011density}. In order to do so, it is crucial that the infinite-temperature TFD state is a product state in the real-space representation, [cf. Eq.~\eqref{eq:product_infinite}] and that, more generally, it maintains low spatial entanglement at finite values of $\beta$. Indeed, this allows us to represent it efficiently as a Matrix Product State (MPS)~\cite{perez2006matrix} with small bond dimension, and apply standard algorithms for its time evolution. 

Unfortunately, the linear growth of the bipartite entanglement entropy poses practical limitations on the time scales which can be simulated~\cite{schollwock2011density}, and our numerical results are often plagued by severe finite-time effects.  For this reason, we limit ourselves to test the quasiparticle prediction in a simple setting where the latter are easier to take into account: we compute the growth of the TFD entanglement entropy for a bipartition of an infinite spin chain. In order to simulate the system directly in the thermodynamic limit we employ the iTEBD algorithm~\cite{vidal2007classical}.

Importantly, while numerical calculations are performed for infinite systems, in the iTEBD algorithm an approximation is made by introducing a finite \emph{bond dimension}~\cite{schollwock2011density} traditionally denoted by $\chi$, which sets an effective cutoff for the maximum entanglement values which can be simulated, and allows us to estimate the interval of validity of our numerical computations.

From the point of view of the quasiparticle picture, Eq.~\eqref{eq:quasiparticle} simplifies when considering half of an infinite system, since there is no saturation of entanglement. Specifically, taking $\ell\to\infty$ in~\eqref{eq:quasiparticle}, we obtain 
\be\label{eq:quasiparticle_growth}
S(t) =  2t \sum_n \int \mathrm{~d} \lambda|v_n(\lambda)| s_n(\lambda)\,,
\ee
where the integral is over all the allowed quasimomenta. This equation is expected to be exact in the limit $t\to\infty$, and yields a prediction for the asymptotic rate of growth $dS(t)/dt$ of the bipartite entanglement entropy.

As before, in order to evaluate~\eqref{eq:quasiparticle_growth}, one needs to specify the quasiparticle content of the model, i.e. the types of possible quasiparticles, together with their velocity and entropy contribution at thermal equilibrium. In the XXZ Heisenberg model, these data can be obtained via the so-called Thermodynamic Bethe Ansatz~\cite{takahashi2005thermodynamics}. Contrary to the non-interacting case, they don't have an elementary form and their structure depends on the value of $\Delta$. Therefore, in the following, we will consider separately the regimes $\Delta>1$ and $0<\Delta<1$ , and test the quasiparticle prediction in the two cases~\cite{Note2}.

\subsection{The regime $\Delta>1$}

The structure of the quasiparticles is particularly simple when $\Delta>1$. In this case, quasiparticles can form bound states of arbitrary numbers, and, in the thermodynamic limit, the system is characterized by an infinite set of \emph{rapidity distribution functions} $\{\rho_n(\lambda)\}_{n=1}^{\infty}$. Here $\lambda\in [-\pi/2,\pi/2]$, while the integer $n$ indicates the number of quasiparticles in the bound states. The functions $\rho_n(\lambda)$ can be thought of as a generalization of the quasimomentum occupation numbers describing non-interacting quantum gases at thermal equilibrium. Together with $\{\rho_n(\lambda)\}$, the state of the system is also characterized by the set of \emph{hole} distribution functions $\{\rho^{h}_{n}(\lambda)\}$, giving information on the density of vacant quasimomenta which could be occupied by the quasiparticles.
 
Due to interactions, the relation between $\{\rho_n(\lambda)\}$ and $\{\rho^{h}_{n}(\lambda)\}$ is non-trivial, and takes the form~\cite{takahashi2005thermodynamics}
\be
\rho_{m}(\lambda)+\rho_{m}^{h}(\lambda)=a_{m}(\lambda)-\sum_{n=1}^{\infty}\left[a_{m n} * \rho_{n}\right](\lambda)\,,
\label{eq:Bethe_equations}
\ee
where we defined
\be
(f \ast g)(\lambda)=\int_{-\pi / 2}^{\pi / 2} d \mu f(\lambda-\mu) g(\mu)
\ee
and
\bea
a_{m n}(\lambda)&=&\left(1-\delta_{m n}\right) a_{|m-n|}(\lambda)+2 a_{|m-n|+2}(\lambda)+\ldots+2 a_{m+n-2}(\lambda)+a_{m+n}(\lambda)\,,\\
a_{n}(\lambda)&=&\frac{1}{\pi} \frac{\sinh (n \eta)}{\cosh (n \eta)-\cos (2 \lambda)}\,.
\eea
with $\cosh(\eta)=\Delta$. 

The solution to~\eqref{eq:Bethe_equations} is not unique, and an additional set of equations must be provided in order to completely determine the quasiparticle distribution functions at inverse temperature $\beta$. Introducing the function 
\be\label{eq:eta_def}
\eta_n(\lambda)=\frac{\rho^h_n(\lambda)}{\rho_n(\lambda)}\,,
\ee
the latter take the form~\cite{takahashi2005thermodynamics}
\be\label{eq:TBA}
\log \eta_{n}(\lambda)=-\beta \pi \sinh(\eta) a_n(\lambda)+\sum_{m=1}^{\infty}\left[a_{n m} * \log \left(1+\eta_{m}^{-1}\right)\right](\lambda)\,.
\ee

Eqs.~\eqref{eq:Bethe_equations} and \eqref{eq:TBA} are a closed set of equations, and can be solved by standard iterative methods. Clearly, in order to do so, one must truncate the infinite system, keeping only a finite number $n_{\rm max}$ of equations (with the accuracy of the numerical solution increasing with $n_{\rm max}$). The functions $\rho_n(\lambda)$ and $\rho_n^{h}(\lambda)$ obtained in this way completely determine the thermal properties of the system. In particular, they allow one to obtain the velocities of the quasiparticles $v_n(\lambda)$ as the solution to~\cite{bonnes2014light}
\be
[\rho^h_{n}(\lambda)+\rho_{n}(\lambda)]v_n(\lambda)=-\frac{\sinh\eta}{2}a^\prime_{m}(\lambda)-\sum_{n=1}^{\infty}\left[a_{m n} * \rho_{n}v_n\right](\lambda)\,,
\label{eq:velocities}
\ee
while $s_{n}(\lambda)$ is given by the so-called Yang-Yang entropy~\cite{takahashi2005thermodynamics}
\be\label{eq:yang_yang}
s_n(\lambda)=[\rho_{n}(\lambda) +\rho^h_{n}(\lambda)]\ln[\rho_{n}(\lambda) +\rho^h_{n}(\lambda)]-\rho_{n}(\lambda) \ln \rho_{n}(\lambda)-\rho^h_{n}(\lambda) \ln \rho^h_{n}(\lambda)\,.
\ee

\begin{figure}
	\centering
	\includegraphics[scale=0.4]{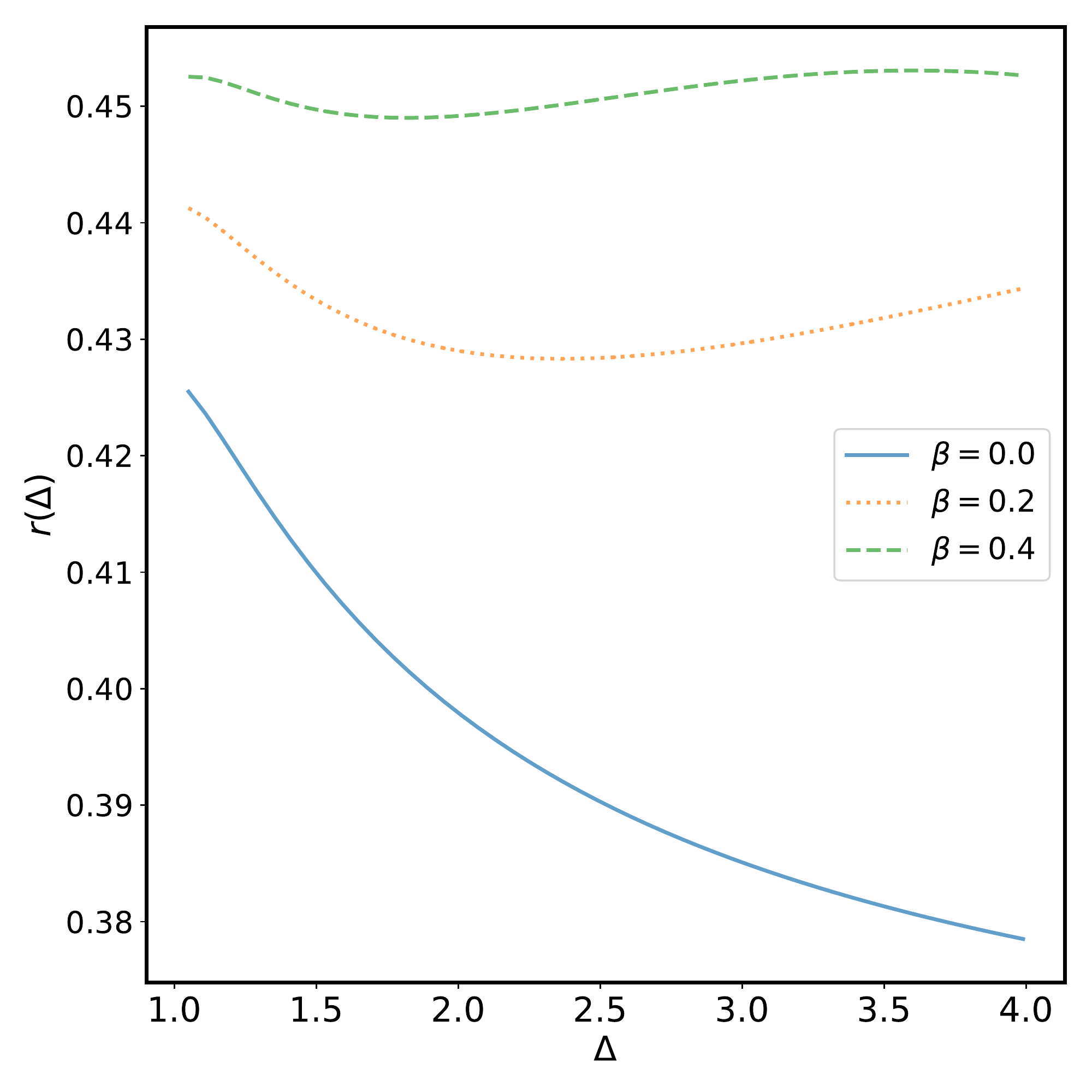}
	\caption{Quasiparticle prediction for the asymptotic rate of growth $r(\Delta)=\lim_{t\to\infty} dS(t)/dt$, as a function of $\Delta>1$ and for different values of $\beta=0,0.2, 0.4$. The curves are obtained by evaluating~\eqref{eq:quasiparticle_growth}, after solving numerically Eqs.~\eqref{eq:Bethe_equations}, \eqref{eq:TBA},~\eqref{eq:velocities}, and using~\eqref{eq:yang_yang}. }
	\label{fig:rampvsDelta_gap}
\end{figure}

We have solved numerically Eqs.~\eqref{eq:Bethe_equations}, \eqref{eq:TBA}, and~\eqref{eq:velocities}. Plugging the result for $v_n(\lambda)$ and $s_n(\lambda)$ [obtained from~\eqref{eq:yang_yang}] into~\eqref{eq:quasiparticle_growth}, we obtain the final quasiparticle prediction for the late-time rate of growth of the bipartite entanglement entropy. A plot for different values of $\beta$ is displayed in Fig.~\ref{fig:rampvsDelta_gap}.

\begin{figure}
	\centering
	\includegraphics[scale=0.475]{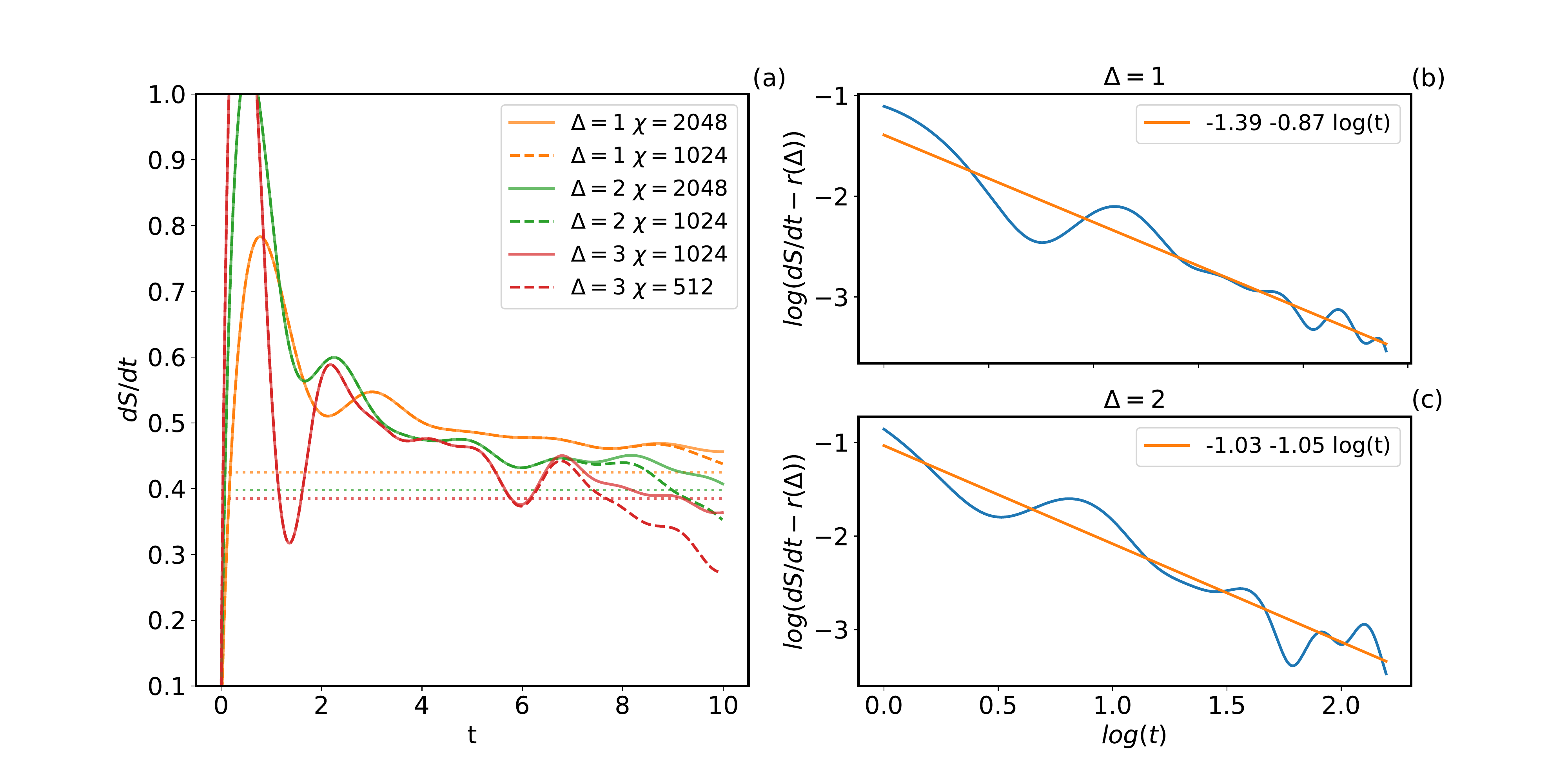}
	\caption{Infinite-temperature TFD entanglement dynamics for $\beta=0$ and $\Delta>1$. Left: Rate of entanglement growth for different value of $\Delta$ as a function of time. Dashed and solid lines correspond to iTEBD data obtained using bond dimension $\chi=1024$ and $\chi=2048$, respectively. Straight dotted lines are the predictions from the quasiparticle picture. Right: Difference between the iTEBD data and the asymptotic formula~\eqref{eq:quasiparticle_growth}, in log-log plot. The straight line is a linear fit. The plot clearly reveals a power-law approach to zero, confirming the validity of the quasiparticle prediction.}
	\label{fig:tfd_gapped_1}
\end{figure}

In Fig.~\ref{fig:tfd_gapped_1} we compare the quasiparticle prediction with our numerical iTEBD calculations. In the first plot, we show $dS(t)/dt$ as a function of time for $\Delta=1,2,3$ at $\beta=0$, reporting data for two different bond dimensions (dashed and solid lines): the time $t_0$ at which the two curves start to deviate from one another gives us an estimate of the maximum time interval for which our data are reliable. We can see that $t_0$ decreases with $\Delta$, which is consistent with the fact that the entanglement entropy appears to grow faster for larger $\Delta$ at short times.

For all the values of the anisotropy, we see short-time oscillations whose amplitude is large compared to the range of variation of $dS(t)/dt$ displayed in Fig.~\ref{fig:rampvsDelta_gap}. This short-time regime is followed by a slow decaying behavior, suggesting large finite-time effects. In the right panel of Fig.~\ref{fig:tfd_gapped_1}, we have plotted the difference between the quasiparticle prediction at late times and $dS(t)/dt$, from which we clearly see a power law decay to zero, finally confirming the validity of our analytic formula~\eqref{eq:quasiparticle_growth}. 

From Fig.~\ref{fig:rampvsDelta_gap}, we see a very weak dependence of the asymptotic value of $dS(t)/dt$ on the inverse temperature $\beta$. On the other hand, the entanglement of the zero-time TFD state increases with $\beta$, further limiting the time scales accessible to our simulations. For this reason, we were not able to perform a meaningful test of the dependence on $\beta$ for $\Delta>1$.

\subsection{The regime $0<\Delta<1$}

The quasiparticle content is significantly more complicated for $\Delta<1$. Setting $\gamma={\rm arccos}(\Delta)$, simplifications occur when $\gamma/\pi$ is a rational number. In this case, quasiparticles can still form bound states, but containing at most a finite number $N_{b}$ of them. The value of $N_b$ and the properties of the quasiparticles depend on the length of the continued-fraction representation of $\gamma/\pi$~\cite{takahashi2005thermodynamics}. In this section, we will restrict ourselves to the case where the latter is at most two. Values of $\gamma/\pi$ with longer continued-fraction representations correspond to a more involved quasiparticle structure, and are only expected to lead to technical, rather than conceptual, complications. We will consider in particular
\be\label{eq:gamma_1}
\gamma=\frac{\pi}{1+\nu_1}\,,
\ee
with $\nu_1=2,3,\ldots$, where we have $N_b=1+\nu_1$, and also
\be\label{eq:gamma_2}
\gamma=\frac{\pi}{\nu_1+\frac{1}{\nu_2}}\,,
\ee
with $\nu_1, \nu_2\geq 2$. In this case, the maximum number of quasiparticles forming a bound state is $N_b=\nu_1+\nu_2$.

As before, we can characterize the system in the thermodynamic limit in terms of the rapidity and hole distribution functions $\{\rho_n(\lambda)\}$, $\{\rho^{h}_n(\lambda)\}$, where now $n=1,\ldots N_b$, and $\lambda\in(-\infty, +\infty)$. The relation between the two reads
\be\label{eq:BE_gapless}
a_{j}(\lambda)=\sigma_{j}\left[\rho_{j}(\lambda)+\rho_{j}^{h}(\lambda)\right]+\sum_{k=1}^{N_b} [a_{j k}\ast \rho_{k}](\lambda)\,,
\ee
where
\be
(f \ast g)(\lambda)=\int_{-\infty}^{\infty} d \mu f(\lambda-\mu) g(\mu)\,,
\ee
and
\begin{align}
a_{j k}(\lambda) &=\left(1-\delta_{n_{j} n_{k}}\right) a_{\left|n_{j}-n_{k}\right|}^{v_{j} v_{k}}(\lambda)+2 a_{\left|n_{j}-n_{k}\right|+2}^{v_{j} v_{k}}(\lambda)+\ldots+2 a_{n_{j}+n_{k}-2}^{v_{j} v_{k}}(\lambda)+a_{n_{j}+n_{k}}^{v_{j} v_{k}}(\lambda)\,,\label{eq:a_j_gapless}\\
 a_{n_{j}}^{v_{j}}(\lambda) &=\frac{v_{j}}{\pi} \frac{\sin \left(\gamma n_{j}\right)}{\cosh (2 \lambda)-v_{j} \cos \left(\gamma n_{j}\right)}\,,\\
 \sigma_{j}&=\operatorname{sign}\left(q_{j}\right)\,.\label{eq:sigma}
\end{align}
In Eqs.~\eqref{eq:a_j_gapless},~\eqref{eq:sigma}, $n_j$, $v_j$ and , $q_j$ depend on the length of the continued-fraction representation of $\gamma$. In the case~\eqref{eq:gamma_1}, we have
\be
\left\{
\begin{array}{ll}
	n_{j}=j, & v_{j}=1, \quad j=1,2 \ldots, \nu_1\,, \\
	n_{\nu_1+1}=1, & v_{\nu_1+1}=-1\,,
\end{array}
\right.
\ee
and
\be
\left\{
\begin{aligned}
	&q_{j}=\nu_1+1-n_{j} \quad j=1,2 \ldots, \nu_1\,, \\
	&q_{\nu_1+1}=-1\,.
\end{aligned}
\right.
\ee
In the case~\eqref{eq:gamma_2}, instead, we have
\begin{align}
n_{j}&= 
\begin{cases}j & 1 \leq j \leq \nu_{1}-1 \,,\\ 
	1+\left(j-\nu_{1}\right) \nu_{1} & \nu_{1} \leq j \leq \nu_{1}+\nu_{2}-1 \,,\\
	 \nu_{1} & j=\nu_{1}+\nu_{2}\,,
 \end{cases}\\
v_{j}&= 
\begin{cases} 
	  +1 & 1 \leq j \leq \nu_{1}-1 \,,\\
	 -1 & j=\nu_{1} \\
	 \exp \left(i \pi \operatorname{floor}\left[\left(n_{j}-1\right) \frac{\nu_{2}}{1+\nu_{1}\,, \nu_{2}}\right]\right) & \nu_{1}+1 \leq j \leq \nu_{1}+\nu_{2}\,,
 \end{cases}
\end{align}
where ${\rm floor}(x)$ is the floor function, and
\be
q_{j}= 
\begin{cases}\frac{1+\nu_{1} \nu_{2}}{\nu_{2}}-j & 1 \leq j \leq \nu_{1}-1 \,,\\ \frac{1}{\nu_{2}}\left(j-\nu_{1}\right)-1 & \nu_{1} \leq j \leq \nu_{1}+\nu_{2}-1\,, \\ \frac{1}{\nu_{2}} & j=\nu_{1}+\nu_{2}\,.
\end{cases}
\ee

Analogously to the case $\Delta>1$, Eq.~\eqref{eq:BE_gapless} has to be complemented with an additional set of equations to determine the thermal stationary state. For both cases~\eqref{eq:gamma_1} and~\eqref{eq:gamma_2}, it reads~\cite{takahashi2005thermodynamics}
\be
\ln \eta_{j}(\lambda)=\beta e_{j}(\lambda)+\sum_{k=1}^{N_b} \sigma_{k} [a_{j k}\ast \ln \left(1+\eta_{k}^{-1}\right)](\lambda)\,,
\label{eq:TBA_gapless}
\ee
where
\be
e_j(\lambda)=-\pi \sin\gamma a_{n_j}^{v_j}(\lambda)\,,
\ee
and $\eta_n(\lambda)$ is defined in~\eqref{eq:eta_def}. Finally, from the knowledge of the thermal rapidity and hole distribution functions, the entropy contribution is given by the Yang-Yang entropy~\eqref{eq:yang_yang}, while the velocity is obtained by solving the system
\be\label{eq:vel_gapless}
\frac{e^{\prime}_j(\lambda)}{2\pi}=\sigma_{j}v_j(\lambda)\left[\rho_{j}(\lambda)+\rho_{j}^{h}(\lambda)\right]+\sum_{k=1}^{N_b} [a_{j k}\ast v_k\rho_{k}](\lambda)\,.
\ee
We have solved numerically Eqs.~\eqref{eq:BE_gapless},~\eqref{eq:TBA_gapless} and \eqref{eq:vel_gapless} by standard iterative methods, after introducing a finite cutoff $\Lambda$ in the space of rapidities.

\begin{figure}
	\centering
	\includegraphics[scale=0.4]{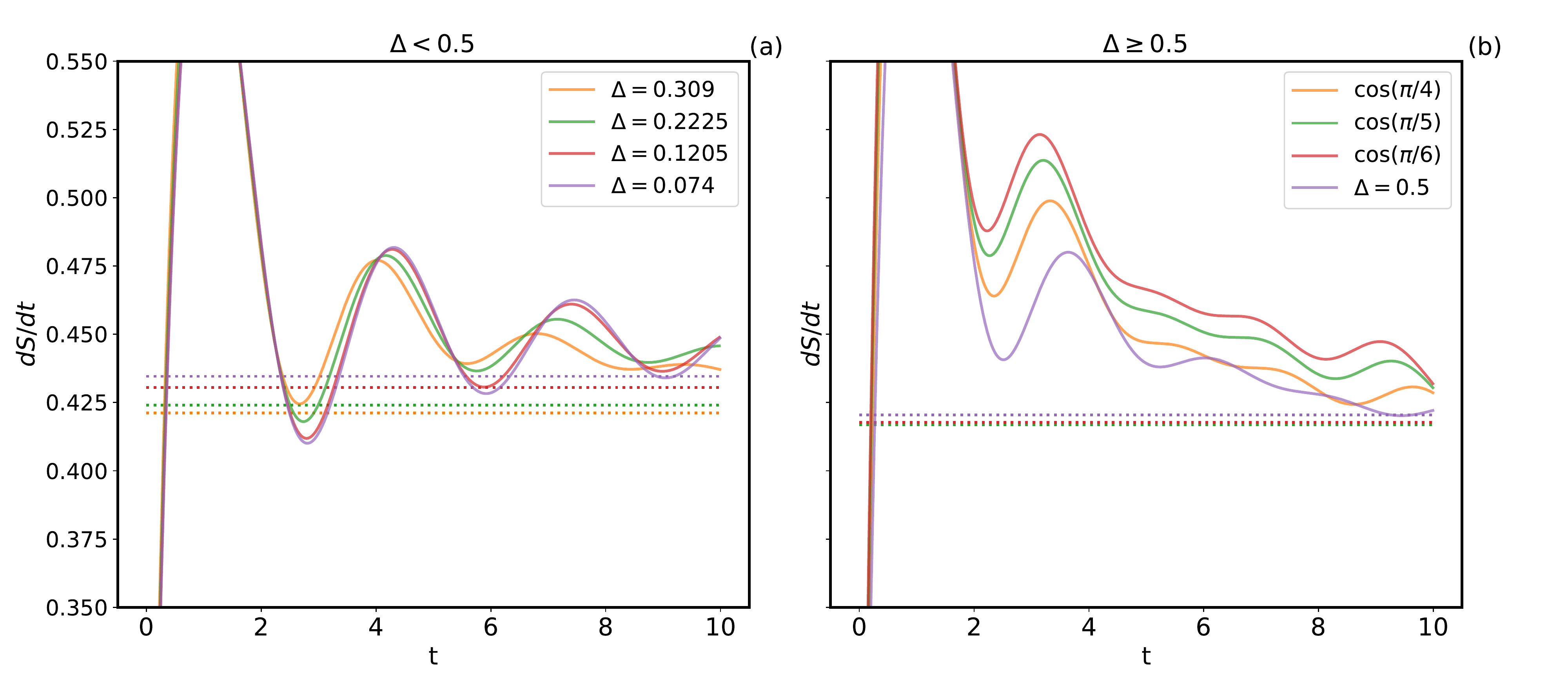}
	\caption{Infinite-temperature TFD entanglement dynamics for $\beta=0$ and $0<\Delta<1$. Solid lines correspond to iTEBD data obtained using bond dimension $\chi=2048$, while straight dotted lines are the predictions from the quasiparticle picture.}
	\label{fig:rate_gapless}
\end{figure}

\begin{figure}
	\centering
	\includegraphics[scale=0.4]{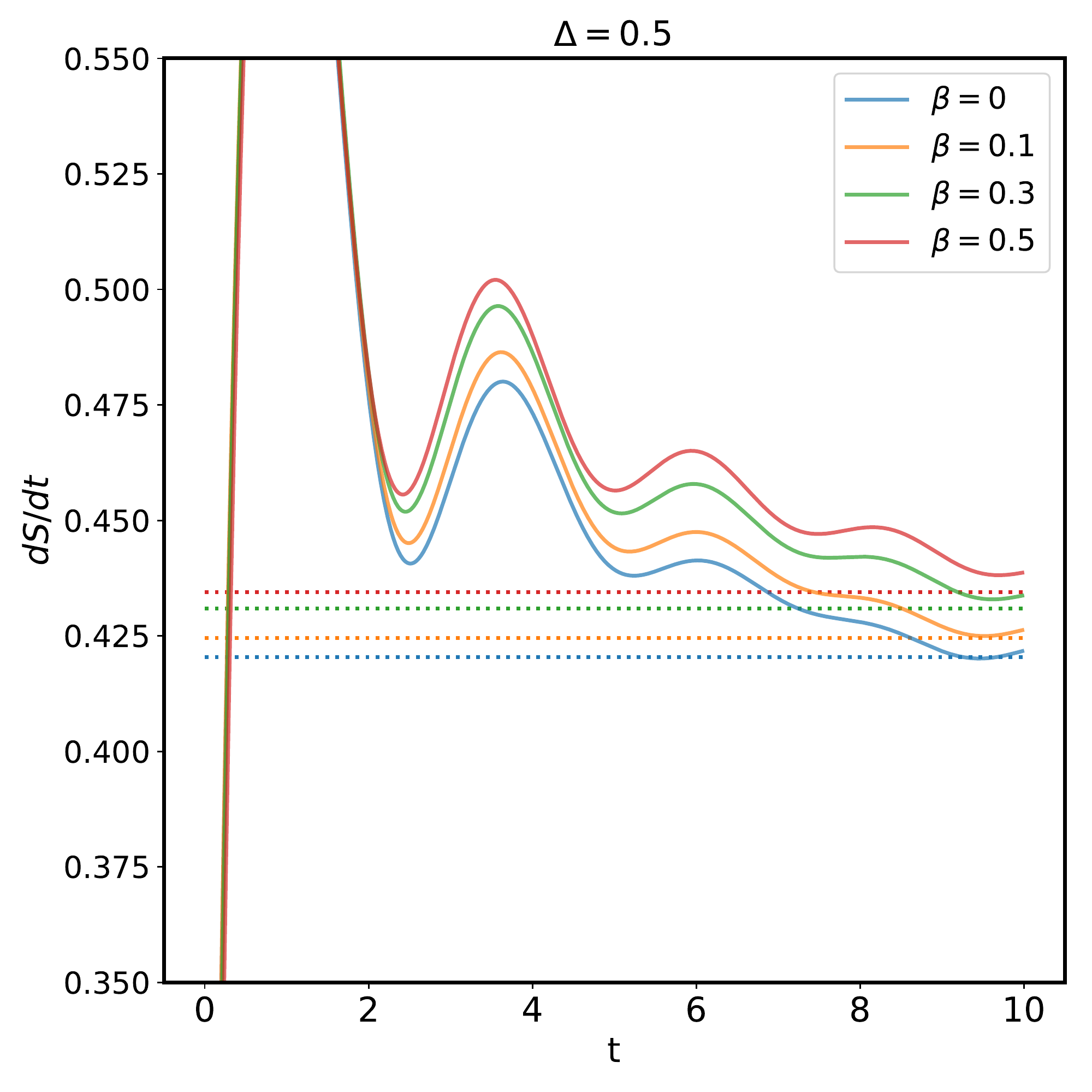}
	\caption{Finite-temperature TFD entanglement dynamics for $\Delta=0.5$. Solid lines correspond to iTEBD data obtained using bond dimension $\chi=2048$, while straight dotted lines are the predictions from the quasiparticle picture.}
	\label{fig:rate_gapless_temperature}
\end{figure}

In Fig.~\ref{fig:rate_gapless} we compare the prediction obtained using Eq.~\eqref{eq:quasiparticle_growth} with our numerical iTEBD calculations. In general, we find that larger finite-time effects are visible, compared to the case $\Delta>1$. For $0.5\leq \Delta<1$, we see that $dS(t)/dt$ appears to decrease in time towards our asymptotic predictions, decaying more slowly as $\Delta$ approaches $1$. Unfortunately, however, although the data are qualitatively consistent with our prediction, the time scales which we can simulate do not allow us to make a more accurate quantitative comparison. 

The situation appears to be worse in the case $\Delta<0.5$ shown in the left panel of Fig.~\ref{fig:rate_gapless}. Here we see large oscillations for the time scales which we can simulate. Perhaps unsurprisingly, oscillations are larger as $\Delta$ approaches zero, and, within the accessible time scales, only for $\Delta= 0.309$ we are able to see hints of an eventual damping in their amplitudes and the onset of a slowly decaying behavior, following this transient short-time regime. In general, we interpret the visible discrepancies as a manifestation of large finite-time effects, due to proximity to the non-interacting point $\Delta=0$. From Fig.~\ref{fig:rate_gapless}, we also see that finite-time effects are less severe for $\Delta=0.5$. For this value of the anisotropy, we are able to test the dependence of the TFD entanglement entropy on $\beta$. Our results are reported in Fig.~\ref{fig:rate_gapless_temperature}, displaying a good quantitative agreement.

\section{Outlook}
\label{sec:conclusions}

We have studied the entanglement dynamics of TFD states in interacting integrable systems. We have shown that the TFD evolution may be interpreted as a quantum quench from an initial state which is low-entangled in the real-space representation and displays a simple quasiparticle structure. Based on these considerations, we have generalized the quasiparticle picture developed for standard quenches and conjectured a formula for the evolution of the von Neumann entanglement entropy, which is expected to be exact in the scaling limit of large sizes and times. 

In the case of integrable spin chains, where exact or efficient numerical calculations can be performed, we have tested our conjecture finding convincing agreement. More generally, our formula applies to continuous quantum field theories, including, for instance, the Lieb-Liniger gas or the relativistic sinh- and sine-Gordon models. Although in these cases it would be difficult to simulate exactly the dynamics, the quasiparticle structure of these models is well known, leading to a straightforward application of our formulas.

As mentioned, the entanglement dynamics of the infinite-temperature TFD state coincides with the operator-space entanglement entropy of the evolution operator~\cite{zanardi2001entanglement,d-17}. Therefore, as a byproduct of our work, we obtained an analytic expression for the latter in generic interacting integrable systems. In this respect, it would be interesting to understand whether a similar quasiparticle description could be obtained for other entanglement-related properties of the evolution operator, 
such as the  tripartite mutual information~\cite{hosur2016chaos}, which characterizes the chaotic and scrambling behavior of the many-body dynamics, 
the negativity \cite{alba2019quantum,ghasemi2021odd,murciano2021quench} or the symmetry resolved entropies \cite{pbc-20,pbc-21}. We leave these questions for future work. 

\section*{Acknowledgments}
P.C. and G.L. acknowledge support from ERC under Consolidator grant number 771536 (NEMO).

\appendix

\section{The TFD state and the Jordan-Wigner transformation}
\label{sec:TFD_JW}

In this section we provide more detail about the JW transformation in the TFD setting. For concreteness, let us assume $N$ even, and define the following left and right Majorana operators
\begin{align}
\tilde\chi^{\alpha}_{2j}&=\left(\bigotimes_{k=1}^{j-1}\sigma_k^{z, \alpha}\right)\sigma_j^{x,\alpha}\,,\qquad  \tilde\chi^{\alpha}_{2j+1}=\left(\bigotimes_{k=1}^{j-1}\sigma_k^{z, \alpha}\right)\sigma_j^{y,\alpha}\,,
\end{align}
where $\alpha=L,R$. This definition corresponds to applying a JW transformation independently on the two replica spaces. We note that the set $\{\tilde \chi^{L}_{j}, \tilde \chi^{R}_{j}\}$ does not satisfy a fermionic algebra, but a mixed one, since
\be
[\tilde \chi^{L}_{j}, \tilde\chi^{R}_{k}]=0\,,\qquad \{\tilde\chi^{\alpha}_{j}, \tilde\chi^{\alpha}_{k}\}=\delta_{j,k}\,.
\ee
In order to obtain a truly fermionic algebra, we introduce
\be
\chi^L_k=i Q^L \tilde\chi^L_k\,,\qquad  \chi^R_k=Q^L \tilde\chi^R_k\,,
\ee
where
\be
Q^L=\prod_{j=1}^{2N} \tilde\chi^L_j\,.
\ee
One can easily verify that $\{\chi_j^{\alpha}\}$ satisfy fermionic anticommutation relations, namely
\be
\{\chi_j^{\alpha},\chi_k^{\beta}\}=\delta_{j,k}\delta_{\alpha,\beta}\,,
\ee
and that $\chi_j^{\alpha\dagger}=\chi_j^{\alpha}$. Furthermore, since $(Q^L)^2=\openone$, we have
\be\label{eq:product}
\prod_{j=1}^s\tilde{\chi}^{\alpha}_{m_j}=\prod_{j=1}^s\chi^{\alpha}_{m_j}\,,
\ee 
for any integer $s$. 

From Eq.~\eqref{eq:product}, we see that we can reformulate the TFD dynamics in terms of the fermionic operators $\chi^{\alpha}_j$. In order to show that the mapping is well defined, however, we also need to make sure that the infinite-temperature state $\ket{\mathcal{I}}$ originally written in terms of spin degrees of freedom, is transformed into a maximally entangled state $\ket{\mathcal{I}_f}$, which is a product state with respect to the fermionic degrees of freedom. To see this, we first note that the state $\ket{\mathcal{I}}$ is completely determined  by the set of relations
\be
(\mathcal{O}^{L}\otimes \openone )\ket{\mathcal{I}}= (\openone \otimes \mathcal{O}^{R, T} ) \ket{ \mathcal I}\,,
\ee
where $\mathcal{O}$ is any operator and $(\cdot)^{T}$ denotes transposition with respect to the computational basis. By rewriting this equation in the Majorana basis we obtain
\be
\chi^{L}_{2k}\ket{\mathcal{I}_f}=-i\chi^{R}_{2k}\ket{\mathcal{I}_f}\,,\qquad  \chi^{L}_{2k+1}\ket{\mathcal{I}_f}=i\chi^{R}_{2k+1}\ket{\mathcal{I}_f}\,,
\ee
i.e. $\ket{\mathcal{I}_f}$ is a purification of the infinite-temperature state, which is also a product state with respect to the fermionic degrees of freedom.

\section{Details on the TFD entanglement dynamics for quadratic fermionic Hamiltonians}
\label{sec:details_TFD_free}

In this Appendix we provide further details about the calculation of the TFD entanglement dynamics for the quadratic Hamiltonian~\eqref{eq:quadratic_fermion}.

We begin by reviewing the diagonalization of the Hamiltonian~\eqref{eq:quadratic_fermion}, which is standard and carried out in two steps. First, we take the Fourier transform of the fermionic operators
\be
c_{j}=\frac{e^{i \pi / 4}}{\sqrt{N}} \sum_{q=0}^{N-1} e^{i \frac{2 \pi}{N} q j} \tilde c_{q}, \qquad \tilde c_{q} \equiv \frac{e^{-i \pi / 4}}{\sqrt{N}} \sum_{j=1}^{N} e^{-i \frac{2 \pi}{N} q j} c_{j}\,.
\ee
Next, we perform a Bogoliubov rotation
\be
\left(\begin{array}{c}
	\tilde{c}_{+q} \\
	\tilde c_{-q}^{\dagger}
\end{array}\right)=\left(\begin{array}{cc}
	\cos \vartheta_{q} & \sin \vartheta_{q} \\
	-\sin \vartheta_{q} & \cos \vartheta_{q}
\end{array}\right)\left(\begin{array}{c}
	b_{+q} \\
	b_{-q}^{\dagger}
\end{array}\right)\,,
\qquad 
\left(\begin{array}{c}
	b_{+q} \\
	b_{-q}^{\dagger}
\end{array}\right)=\left(\begin{array}{cc}
	\cos \vartheta_{q} & -\sin \vartheta_{q} \\
	\sin \vartheta_{q} & \cos \vartheta_{q}
\end{array}\right)\left(\begin{array}{c}
	\tilde c_{+ q} \\
	\tilde c_{-q}^{\dagger}
\end{array}\right)\,,
\ee
where we defined
\be
\tan \left(2 \vartheta_{q}\right)=\frac{\gamma \sin \left(\frac{2 \pi}{N} q\right)}{h-\cos \left(\frac{2 \pi}{N} q\right)}\,.
\ee
The Hamiltonian becomes diagonal in terms of the new modes
\be\label{eq:diagonal}
H= \sum_{q =0}^{N-1} \varepsilon\left(\frac{2 \pi}{N} q\right)\left\{b_{q}^{\dagger} b_{q}-\frac{N}{2}\right\}\,,
\ee
where
\be
\varepsilon(\alpha) \equiv \sqrt{(h-\cos \alpha)^{2}+\gamma^{2} \sin ^{2} \alpha}\,.
\ee

Concatenating the Fourier and Bogoliubov transformation, we may directly write down the initial Majorana modes, defined in Eq.~\eqref{eq:initial_majorana}, in terms of the final ones, defined in Eq.~\eqref{eq:majoranas_k}. We have
\begin{align}
\chi_{j,1}&=\frac{1}{\sqrt{2}}(c_j+c^\dagger_j)=\frac{1}{\sqrt{2}}\sum_{q=0}^{N-1}\left[\frac{1}{\sqrt{N}}e^{i\pi/4}e^{2\pi qj/N}\tilde{c}_q+\frac{1}{\sqrt{N}}e^{-i\pi/4}e^{-2\pi qj/N}\tilde{c}^{\dagger}_q\right]\nonumber\\
&=\frac{1}{\sqrt{2N}}\sum_{q=0}^{N-1}\left[\cos(\vartheta_q)\left(e^{i\pi/4+2\pi q j/N }b_q+ e^{-i\pi/4-2\pi q j/N }b^{\dagger}_q\right)\right.\nonumber\\
&+\left.
\sin(\vartheta_q)\left(e^{i\pi/4+2\pi q j/N }b^{\dagger}_{-q}+ e^{-i\pi/4-2\pi q j/N }b_{-q}\right)
\right]\,.
\end{align}
For the second term in the sum, we may rewrite
\begin{align}
\sum_{q=0}^{N-1}\left[
\sin(\vartheta_q)\left(e^{i\pi/4+2\pi q j/N }b^{\dagger}_{-q}+ e^{-i\pi/4-2\pi q j/N }b_{-q}\right)
\right]=\sum_{q=0}^{N-1}\left[
-\sin(\vartheta_q)\left(e^{i\pi/4-2\pi q j/N }b^{\dagger}_{q}+ e^{-i\pi/4+2\pi q j/N }b_{q}\right)
\right]
\end{align}
and so
\begin{align}
	\chi_{j,1}&=\frac{1}{\sqrt{2N}}\sum_{q=0}^{N-1}\left\{b_q\left[ \cos\vartheta_q e^{i\pi/4+2\pi q i j/N}- \sin\vartheta_q e^{-i\pi/4+2\pi q i j/N} \right]+ b^{\dagger}_q\left[ \cos\vartheta_q e^{-i\pi/4-2\pi q i j/N}- \sin\vartheta_q e^{i\pi/4-2\pi q i j/N} \right]\right\}\nonumber\\
	&\qquad \qquad =\frac{1}{\sqrt{N}}\sum_{q=0}^{N-1}\left\{\psi_{q,1}\left[ \cos\vartheta_q \cos\left( \frac{2\pi q j}{N}+\frac{\pi}{4}\right)- \sin\vartheta_q \cos\left( \frac{2\pi q j}{N}-\frac{\pi}{4} \right)\right]\right.\nonumber\\
	&\qquad \qquad \qquad \qquad+\left.\psi_{q,2}\left[ -\cos\vartheta_q \sin\left( \frac{2\pi q j}{N}+\frac{\pi}{4}\right)+ \sin\vartheta_q \sin\left( \frac{2\pi q j}{N}-\frac{\pi}{4} \right)\right]\right\}\,.
\end{align}
Analogously, we have
\begin{align}
	\chi_{j,2}&=\frac{i}{\sqrt{2N}}\sum_{q=0}^{N-1}\left\{b_q\left[ -\sin\vartheta_q e^{-i\pi/4+2\pi q i j/N}- \cos\vartheta_q e^{i\pi/4+2\pi q i j/N} \right]\right.\nonumber\\
	&+ \left.b^{\dagger}_q\left[ \cos\vartheta_q e^{-i\pi/4-2\pi q i j/N}+ \sin\vartheta_q e^{i\pi/4-2\pi q i j/N} \right]\right\}\nonumber\\
	&\qquad \qquad =\frac{1}{\sqrt{N}}\sum_{q=0}^{N-1}\left\{\psi_{q,1}\left[ \sin\vartheta_q \sin\left( \frac{2\pi q j}{N}-\frac{\pi}{4}\right)+ \cos\vartheta_q \sin\left( \frac{2\pi q j}{N}+\frac{\pi}{4} \right)\right]\right.\nonumber\\
	&\qquad \qquad \qquad \qquad+\left.\psi_{q,2}\left[ \sin\vartheta_q \cos\left( \frac{2\pi q j}{N}-\frac{\pi}{4}\right)+ \cos\vartheta_q \cos\left( \frac{2\pi q j}{N}+\frac{\pi}{4} \right)\right]\right\}\,.
\end{align}

Let us now consider the two-replica Hilbert space and denote by $\psi^{L/R}_{j,\alpha}$ the left/right Majorana fermions. We define the ordered vectors
\begin{align}
	\psi &= (\psi_1 \dots \psi_N)\,, \qquad  \psi_k= (\psi_{k,1}^L,\psi_{k,1}^R,\psi_{k,2}^L,\psi_{k,2}^R)\,,\label{eq:ordering_modes}\\
	\chi &= (\chi_1 \dots \chi_N)\,, \qquad  \chi_k= (\chi_{k,1}^L,\chi_{k,1}^R,\chi_{k,2}^L,\chi_{k,2}^R)\,.
\end{align}
We can rewrite
\begin{equation}
	\chi_j= \sum_k A_{j,k} \psi_k\,,
\end{equation}
where $A=\tilde{A}/\sqrt{N}$ and
\begin{align*}
	[\tilde A_{j,k}]_{1,\alpha}&=\!\!\left[\cos(\vartheta_q ) c^{+}\left( \frac{2\pi q j}{N}\right)- \sin(\vartheta_q) c^{-}\left( \frac{2\pi q j}{N}\right), 0,   -\cos(\vartheta_q )s^{+}\left( \frac{2\pi q j}{N}\right)+ \sin(\vartheta_q) s^{-}\left( \frac{2\pi q j}{N} \right), 0\right]\\
	[\tilde A_{j,k}]_{2,\alpha}&=\!\!\left[0, \cos(\vartheta_{-q} ) c^{+}\left( -\frac{2\pi q j}{N}\right)- \sin(\vartheta_{-q}) c^{-}\left( -\frac{2\pi q j}{N}\right), 0,  -\cos(\vartheta_{-q} )s^{+}\left(- \frac{2\pi q j}{N}\right)+ \sin(\vartheta_{-q}) s^{-}\left(- \frac{2\pi q j}{N} \right)\right]\\
	[\tilde A_{j,k}]_{3,\alpha}&= \!\!\left[ \sin(\vartheta_q) s^{-}\left( \frac{2\pi q j}{N}\right)+ \cos(\vartheta_q) s^{+}\left( \frac{2\pi q j}{N}\right) , 0,  \sin(\vartheta_q )c^{-}\left( \frac{2\pi q j}{N}\right)+ \cos(\vartheta_q) c^{+}\left( \frac{2\pi q j}{N} \right)
	, 0\right]\\
	[\tilde A_{j,k}]_{4,\alpha}&= \!\! \left[0,  \sin(\vartheta_{-q}) s^{-}\left(- \frac{2\pi q j}{N}\right)+ \cos(\vartheta_{-q}) s^{+}\left( -\frac{2\pi q j}{N}\right), 0,   \sin(\vartheta_{-q} )c^{-}\left( -\frac{2\pi q j}{N}\right)+ \cos(\vartheta_{-q}) c^{+}\left(- \frac{2\pi q j}{N} \right)\right]
\end{align*}
with
\be
c^{\pm}(x)=\cos(x\pm \pi/4)\,, \qquad s^{\pm}(x)=\sin(x\pm \pi/4)\,.
\ee

Now, the covariance matrix of a TFD state corresponding to the diagonal Hamiltonian~\eqref{eq:diagonal} is immediate to compute. In the basis of the Majorana modes $\psi_j$, ordered as in~\eqref{eq:ordering_modes}, it reads (a similar calculation was performed in Ref.~\cite{jiang2019circuit})
\be
\Omega(t,\beta)=\bigoplus_k \Omega_k(t,\beta)\,,
\ee
with
\begin{equation}
	\Omega_k(t,\beta) = 
	\begin{pmatrix}
		0 & \sin[2 \varphi(k)] \sin[\eps(k) t] & \cos[2 \varphi(k)] & \sin[2 \varphi(k)]  \cos[\eps(k) t] &  \\
		-\sin[2 \varphi(k)] \sin[\eps(k) t] & 0 & -\sin[2 \varphi(k)]  \cos[\eps(k) t] &  \cos[2 \varphi(k)] \\
		-\cos[2 \varphi(k)] & \sin[2 \varphi(k)]  \cos[\eps(k) t] & 0 &  -\sin[2 \varphi(k)] \sin[\eps(k) t] \\
		-\sin[2 \varphi(k)]  \cos[\eps(k) t] & -\cos[2 \varphi(k)] &  \sin[2 \varphi(k)] \sin[\eps(k) t] & 0 
	\end{pmatrix}\,,
\end{equation}
and
\be
\varphi(k)=\arctan( e^{-\beta \eps (k)/2})\,.
\ee
Putting all together, we arrive at the following result for the covariance matrix of the TFD state corresponding to the Hamiltonian~\eqref{eq:quadratic_fermion}:
\be
\Gamma(t,\beta)=A \Omega(t,\beta) A^{T}\,.
\ee

\end{document}